# Quality Evaluation of Projection-Based VR Displays

Dave Pape, Dan Sandin
*Electronic Visualization Laboratory*
*University of Illinois at Chicago*
{pape, dan}@evl.uic.edu

## Abstract

*We present a collection of heuristics and simple tests for evaluating the quality of a projection-based virtual reality display. A typical VR system includes numerous potential sources of error. By understanding the characteristics of a correctly working system, and the types of errors that are likely to occur, users can quickly determine if their display is inaccurate and what components may need correction.*

## Introduction

Typical virtual reality systems include many elements which must be configured and functioning correctly to provide an accurate virtual world simulation. Some of these parts, such as electromagnetic trackers and CRT projection displays, are prone to errors. VR researchers may often be willing to accept some errors, because it's a research system. Real-world end users, however, will expect systems to operate correctly. Corporations such as Caterpillar and General Motors are using VR in prototyping new products (Lehner and DeFanti, 1997; Smith, 1996). They need accurate visuals for this process to be truly useful. In support of this goal, we have assembled a suite of practical tests that may be used to check the quality of a display without requiring extensive time or special equipment.

With a full understanding of how things should function, it is possible to examine the output of a system to determine if everything really is working correctly. The goal of the tests we describe here is to be able to quickly determine whether there is an error, and what the source of that error is likely to be, without a lot of special equipment or complicated procedures. Many of these tests can be performed within any typical application program which uses the VR system. Some of the more precise tests are separate programs which require basic tools, such as tape measure, small targets, or a video camera.

The tests described here are solely concerned with evaluating the visual portion of a VR system, that is, viewer-centered, stereoscopic images. Hence we focus on the parts which contribute to that – the display itself, and the tracking system. We also focus on projection-based displays, although a number of the tests are also applicable to HMD based augmented reality systems.

We will first identify and describe the relevant system components. Then we will review how the system functions, to enumerate various characteristics of a correct display. We will describe the tests that can be made to check an actual system against these ideal characteristics. Finally, we will present the special-purpose tests which provide more in-depth analysis.

## Related Work

In practice, there are a number of common sources of error. Many authors have presented methods of correcting or reducing these errors.

Bryson (1992), Ghazisaedy et al. (1995), and Livingston and State (1997) address static errors in position and orientation data from six degree-of-freedom tracking systems. Each of these corrects the tracking by measuring the errors either creating a lookup table or a polynomial fit function; the method of actually measuring the errors differs in each approach. Bryson (1992) required the least complicated equipment, attaching the tracked sensor to a peg-board that defined a rectilinear grid. Ghazisaedy et al. (1995) used an ultrasonic device to measure the actual position of a sensor and compare it to the reported data. Livingston and State (1997) used a mechanical tracker, a Faro arm, to obtain correct measurements; in this case, both position and orientation data were calibrated.

Other methods have been presented to measure dynamic tracking errors, i.e. tracker latency. Bryson and Fisher (1990) recorded video of experiments, combining the computer output with video of a user moving a tracked device. The recordings were then examined frame-by-frame to determine the time difference between an action and the resulting computer image. Liang et al. (1991) attached a sensor to a gravity-driven pendulum and compared timestamps of observed and tracker detected zero-crossings. Adelstein et al. (1996) used a motor-driven swing arm to move a tracked sensor at different rates, measuring the state of the tracker data and the swing arm with a single PC.

Deering (1992) described a number of issues which must be dealt with to generate precise, accurate images for a head-tracked user on a CRT monitor. Deering and Sowizral (1997) listed steps taken to finely calibrate a larger scale, projection-based display; the projection screens, projectors, and tracking system were all precisely aligned and measured in order to be able to generate exact, viewer-centered renderings.

In the case of projected displays, potential problems such as color balance and image geometry must be dealt with, as described by von Erdmansdorff (1999). Multi-screen systems require these image details to correspond seamlessly across multiple displays. In addition to matching the projectors themselves, the computer generated images must match; Reiners (1999) identified rendering library features that are necessary to maintain inter-screen continuity of effects such as specular highlights and fog.

An important factor in most error analyses and reductions is that they can require a significant amount of time and precise measurement. However, once a VR system is in operation, users would like to be able to check the system for significant errors quickly; in many cases while running an application, if an error is suspected. Once errors have been identified, the more precise correction methods may be applied.

## Projection-Based VR Components

A full VR system can contain many different components. The ones that we are concerned with here are the tracking and stereoscopic display.

A projection-based display system, such as a CAVE®, commonly uses a 6 degree-of-freedom tracking device, with two or more sensors. One sensor tracks the user's head position and orientation, the others track a wand, gloves, or other control devices. The majority of currently fielded systems use electromagnetic trackers.

The typical display device is a CRT projector and rear projection screen. In large-scale systems like CAVEs, there are multiple projectors and screens which are edge-aligned. Stereoscopic display in projection-based systems is most commonly done using liquid crystal shutter glasses, which are synchronized with the frame-sequential stereo video on the screens.

The objective of the system is to produce a viewer-centered, stereoscopic display; each of the user's eyes should see an image correctly rendered from that eye's position through the window of the screen(s). The display pipeline to accomplish this is as follows. The tracking system reports the latest position and orientation of the head and wand trackers. From the head data, and pre-defined offsets and inter-pupillary distance (IPD), the positions of the user's eyes are calculated. For each screen and each eye, the eye position and the pre-defined location of the projection plane (the display screen), is used to calculate an off-axis perspective projection. This off-axis perspective is particularly important, as it is necessary both to provide correct views as the user moves around, and to maintain seamless matching of images across multiple screens (see figure 1). This projection is used to render the world database to the frame buffer. When all rendering has been completed, buffers are swapped to display the result. The shutter glasses mediate the video so that each eye sees only the images that were rendered for it. In a multiple-screen system, all the images must remain synchronized to present the illusion of a single, continuous display. This means swapping all buffers at the same time, and genlocking the video so that the video and stereo phases are the same on all screens.

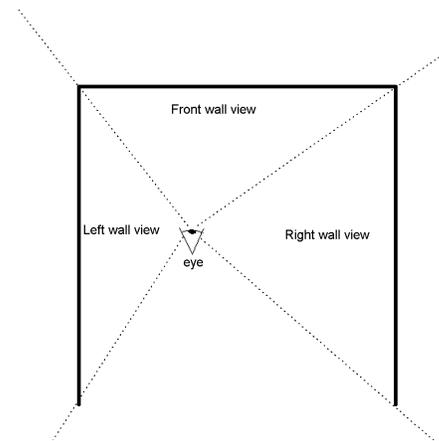

Figure 1. Off-axis perspective in a CAVE

There are several basic errors which might occur in this system. The tracker data may be inaccurate, either in the form of spatial distortion, jitter, or temporal lag. These errors may be due to inherent problems with the tracking technology, environmental factors such as metal objects, or misconfiguration of the hardware. The projected displays may be inaccurate; this can include poor geometric alignment or miscalibration of colors. In a multiple projector display, images from different projectors might not match properly. The stereoscopic display may be faulty; in a CAVE, one screen might be correct while the eye-views are reversed on another screen. Finally, the perspective projection may be incorrect; this can result from tracking errors, incorrect IPD, errors calculating the eye positions from the head data, or incorrect data describing the projection planes.

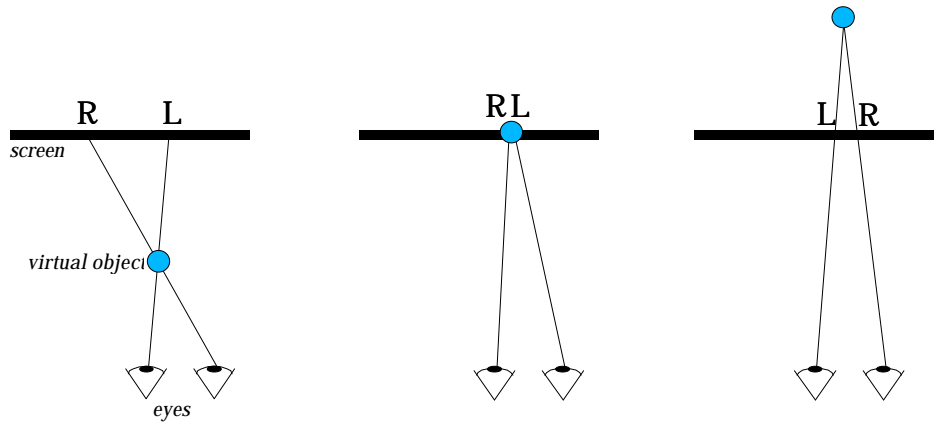

Figure 2.  Screen image location based on object position

## Characteristics of a Correct Display

There are a number of simple properties of the displayed images of virtual objects, which will hold true when everything is working correctly.  Some of them are straightforward, while others can appear counter-intuitive to the casual user (such as when objects 'shrink' as one approaches the screen).  Understanding these properties will help one to better determine whether a system is operating correctly or not.

The behavior of the stereo disparity of a virtual object's image, that is, the displacement between the two rendered images of the object as seen on the screen itself, depends on the object's position relative to the screen (see figure 2). If the object is in front of the screen, on the same side as the user, then the left eye image (L) will appear to the right of the right eye image (R).  If the object is behind the screen, this is reversed – the left eye image appears to the left.  If the object is exactly coincident with the screen, then there is no disparity – the two images appear superimposed.  If an object is effectively at infinity, i.e. extremely far away, the disparity will be equal to the IPD.

The movement of an object's image on the screen in response to the user's head will follow a similar pattern.  For an object in front of the screen, the image will move in the direction opposite that of the head.  For an object positioned on the screen, it will not move.  For an object behind the screen, it will move in the same direction as the head.  If the object is at infinity, it will move the same amount as the head.

As the user's eyes approach the screen, the field of view through that screen – the width of the perspective frustum – increases.  As a result, for objects behind the screen, their projected images appear smaller, in terms of the number of pixels covered on the screen (see figure 3).  This particular effect has been known to confuse many new users of projection-based displays, and can lead them to believe something is wrong with the system when it is in fact behaving correctly.

In the extreme, if the user's eyes are immediately at the plane of the screen, the perspective projection will expand to a half-space.  In this case, the displayed image will appear to "explode"  (see figure 4).

In a CAVE (or, more generally, any system with multiple, non-coplanar screens), straight lines and polygon edges which cross between two screens should always appear straight to the tracked user.  However, when a scene is viewed from any position other than that for which the views were rendered, the perspective will be incorrect.  Then, straight lines will appear to bend at the screen joins (see figure 5).

A few other simple properties include: if a virtual pointer is drawn at the tracked wand position, it should always appear to be at the same position as the physical wand, no matter where the wand or the viewpoint moves.  In a reasonably large, flat world, the horizon line will be nearly level with the user's eyes. The direction of the stereo disparity for an object behind the screen matches the orientation of the head; when the user's eyes are level, the two images should also be level.

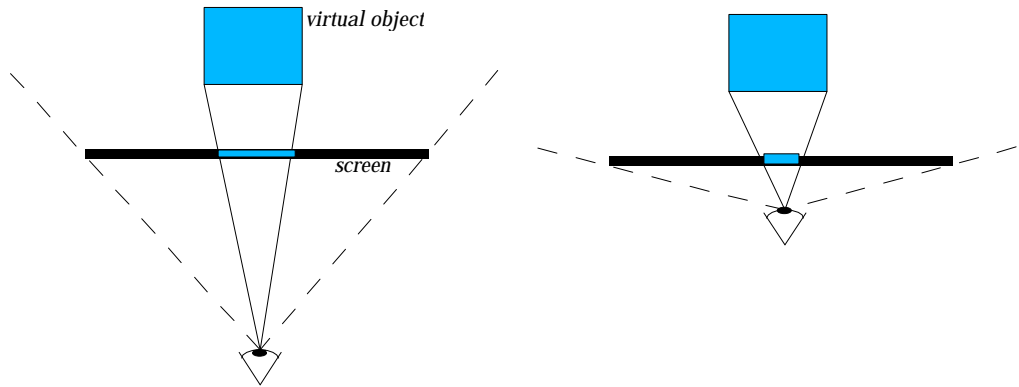

Figure 3. An object appears to shrink as the eyes move toward the screen

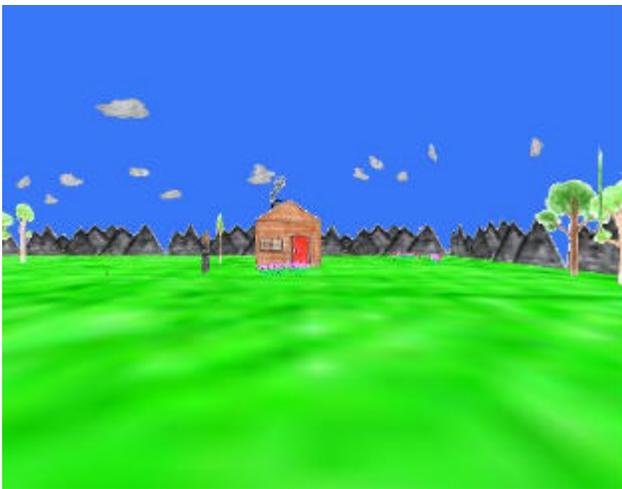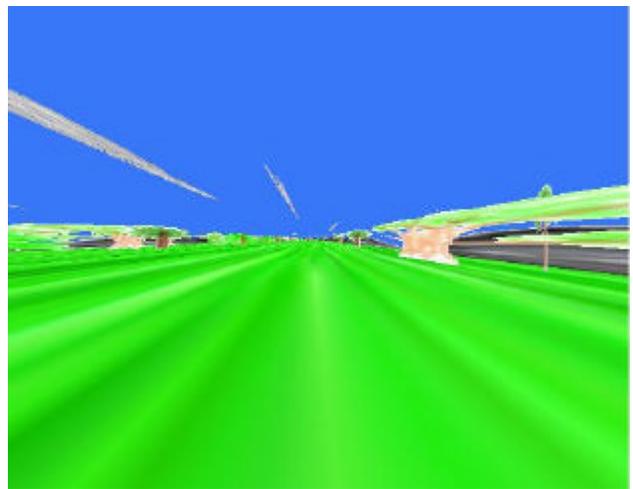

Figure 4. The image "explodes" as the eyes reach the screen

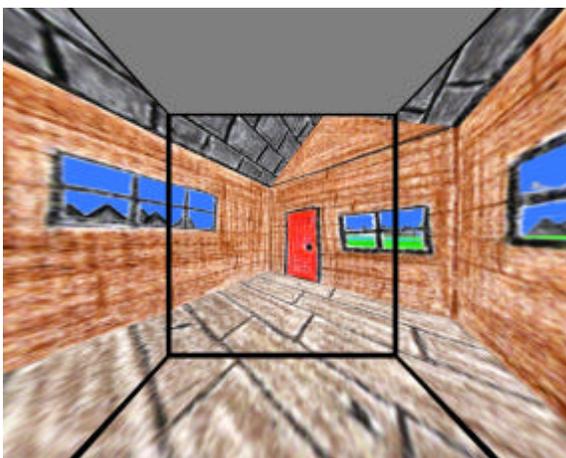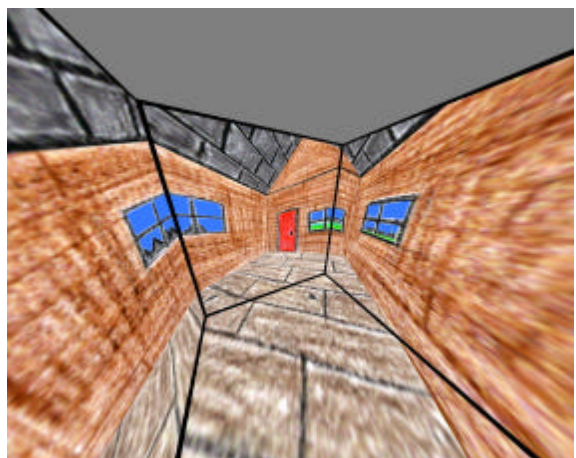

Figure 5. Straight lines appear to bend, when viewed from the wrong position

## Application Tests

While running a VR application, the above characteristics can be tested to determine the general quality and accuracy of the system. These tests can rapidly determine if something is incorrect; however, certain visual errors may, in theory, result from more than one possible cause. Performing several different tests will narrow down the likely cause; the more specific tests described later can isolate the different contributing factors.

The tests that can be performed are:
- orientation of tracker's coordinate system
- tracking position offset
- left/right stereo phase
- location of projection plane
- end-to-end tracking latency
- edge-matching between screens

Many of the display properties identified earlier can be confirmed by removing the tracked glasses and moving them about while observing the resulting images on the screen. Given an object in the virtual world which is known to be located on the far side of the screen, move the glasses around parallel to the screen; if the tracking orientation is correct, the image of the object should move the same way as the glasses. Moving the glasses toward the screen and checking that the displayed image appears to shrink will confirm the remaining axis of the tracking.

If the application has a sufficiently distant horizon line, this can be used to test the vertical component of tracking. Move the glasses up and down, and check that the horizon remains level with them.

In a multi-screen system such as a CAVE, tracking position errors can also be checked by looking at straight edges. Position an object so that one or more edges cross screen boundaries, and look at them through the tracked glasses. If the tracking is accurate, the edges will appear straight. If it is inaccurate, they will appear bent; in that case, the scale of the error can be estimated by removing the tracked glasses and moving them until the edges do look straight (from the now un-tracked viewpoint).

The stereo phase can be checked by looking through the glasses one eye at a time. Using an object whose position in front of or behind the screen is known, observe whether the relative position of the left and right images matches that described earlier.

Moving the glasses toward the screen until the image explodes will determine where the projection plane is, in the tracker's coordinate system. That is, it shows how well the tracking system and screen coordinates are registered to each other. If the application includes objects which can be placed exactly at the presumed location of the screen, this can confirm the projection plane independently of the tracking – when the objects are at the screen location, they should appear "flat"; i.e. the left and right images should be the same.

Overall latency in the tracking and rendering can be judged by observing a virtual object that has its position attached to the wand. Simply moving the wand and seeing how much the object lags behind gives a general sense. An approximation of the actual delay can be determined by waggling the wand back and forth quickly. When the speed of waggling is fast enough that the virtual image's phase is exactly opposite that of the wand – that is, the image is at the left extreme when the wand is at the right extreme and vice versa – the latency is half the period of the waggling motion. Latencies on the order of 100 ms can be easily estimated this way.

In a multi-screen system, placing virtual objects so that their images span multiple screens will check well the screens match each other. The matching of object edges will confirm projector calibration and screen geometry; looking at these edges through the glasses is useful for confirm whether the stereo phase is the same on all walls. Comparing the images on the two screens will check color balance, although details of the rendering may also affect this, as described by Reiners (1999), so one should be aware of whether the application compensates for these problems.

## Standalone Tests

Evaluating a running application is an easy way to determine if everything is working correctly. However, if something is found to be wrong, in some cases there may be more than one possible cause – an image may appear in the wrong spot on the screen because the tracker data is wrong, the projection parameters are wrong, or because the CRT projector is not aligned properly. To more precisely isolate specific problems, we describe here three simple standalone tools.

The first thing that should be verified in a system is the projector alignment and calibration. This is not dependent on the tracking or other parts of the system, and if it is incorrect, tests of the other components may be invalid. We use a custom video test pattern when setting up projectors, and can use this test pattern to check them later. The test pattern includes rectangular gridlines spaced 6" apart. This allows one to verify the linearity of image. It is also useful for matching edges across adjacent screens. There are color bars for matching color between screens. Color-match checking should always be done wearing the stereo glasses, as they affect brightness and color as seen by user. The test pattern also includes stereo bars – two sets of vertical bars, one drawn only for left eye, the other for right eye, drawn at different positions in the frame. This is to verify stereo synchronization of the glasses. The vertical bars stretch from top to bottom to check for genlock problems – if stereo is not genlocked, one part of the image might be correct, but not the frame as a whole. This will appear as a break, where the image switches between the left and right eye views (see figure 6). The test bars also allow one to observe any excessive stereo ghosting that might be caused by slow video phosphors, a problem described by Cruz-Neira et al. (1993).

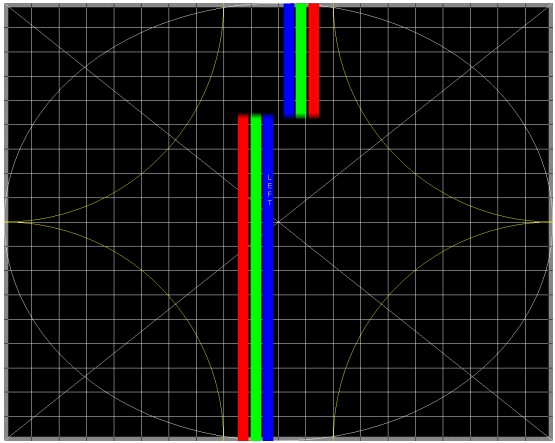

Figure 6. Incorrect stereo genlock produces a break

Next is the basic tracker confidence test. This is a standard CAVE program, so that it reads tracking data and applies all the same transformations as any application. It displays the numeric values of tracker data, which can be verified with a tape measure. This includes the head position, as well as the derived eye positions. The program draws a marker attached to wand, to check how well it stays attached. It also draws a marker at a fixed point in space – 5 feet above the origin on the floor; when the tracking is inaccurate, this marker may give a better idea of the magnitude of the error than the wand marker, since both the head and wand tracking are likely to be affected.

Finally, the line of sight tracking calibration program (Czernuszenko et al., 1998), uses the expected correspondence between real and virtual objects to examine and correct static tracking errors. In this method, physical targets are placed in the CAVE at known locations. The program draws markers at same virtual positions as physical targets. A user can then move around and view the targets from different locations, to see how well the virtual markers match them. The calibration lookup table can then be adjusted dynamically by changing the tracker offset at any location until the image lines up properly with the targets.

## Conclusion

We have listed a set of display characteristics and tests that can be used to evaluate the quality of a VR environment. They derive from the design of projection-based VR systems, but some, such as the latency tests and matching of real and virtual elements, are applicable to other systems such as augmented-reality HMDs. Knowledge of how a display should function, and the likely sources of errors, allows users to quickly check an existing system and spend time on more involved calibration methods only when it is actually necessary, thus improving the general usefulness of these systems.

## Acknowledgments

The virtual reality research, collaborations, and outreach programs at the Electronic Visualization Laboratory (EVL) at the University of Illinois at Chicago are made possible by major funding from the National Science Foundation (NSF), awards EIA-9802090, EIA-9871058, ANI-9980480, ANI-9730202, and ACI-9418068, as well as NSF Partnerships for Advanced Computational Infrastructure (PACI) cooperative agreement ACI-9619019 to the National Computational Science Alliance. EVL also receives major funding from the US Department of Energy (DOE), awards 99ER25388 and 99ER25405, as well as support from the DOE's Accelerated Strategic Computing Initiative (ASCI) Data and Visualization Corridor program. In addition, EVL receives funding from Pacific Interface on behalf of NTT Optical Network Systems Laboratory in Japan.

CAVE is a registered trademark of the Board of Trustees of the University of Illinois.

## References

B. D. Adelstein, E. R. Johnston, and S. R. Ellis. (1996) "Dynamic Response of Electromagnetic Spatial Displacement Trackers." *Presence: Teleoperators and Virtual Environments*, Vol. 5, No. 3. pp. 302-318.

S. Bryson (1992). "Measurement and Calibration of Static Distortion of Position Data from 3D Trackers." *Stereoscopic Displays and Applications III, Proceedings SPIE 1669*, pp. 244-255.

S. Bryson and S. S. Fisher (1990). "Defining, Modeling, and Measuring System Lag in Virtual Environments." *Stereoscopic Displays and Applications I, Proceedings SPIE 1256*, pp. 98-109.

C. Cruz-Neira, D. J. Sandin, and T. A. DeFanti (1993). "Surround-Screen Projection-Based Virtual Reality: The Design and Implementation of the CAVE."

*Proceedings of SIGGRAPH '93 Computer Graphics Conference*, pp. 135-142.

M. Czernuszenko, D. Sandin, T. DeFanti (1998). "Line of Sight Method for Tracker Calibration in Projection-Based VR Systems". *Proceedings of the 2nd International Immersive Projection Technology Workshop*.

M. Deering (1992). "High resolution virtual reality." *Computer Graphics (SIGGRAPH '92 Proceedings)*, Vol. 26 No. 2, pp. 195-202.

M. Deering, and H. Sowizral (1997). "A General View Model for Virtual Reality Displays." *Proceedings of the 1st International Immersive Projection Technology Workshop*, pp. 27-38.

A. von Erdmansdorff (1999). "A Comparison of CRT and LCD Technology". *Proceedings of the 3rd International Immersive Projection Technology Workshop*, pp. 95-106.

M. Ghazisaedy, D. Adamczyk, D. J. Sandin, R. Kenyon, and T. A. DeFanti (1995). "Ultrasonic Calibration of a Magnetic Tracker in a Virtual Reality Space." *Proceedings of the IEEE Virtual Reality Annual International Symposium (VRAIS '95)*.

V. D. Lehner and T. A. DeFanti (1997). "Distributed Virtual Reality: Supporting Remote Collaboration in Vehicle Design." *Computer Graphics and Applications*, Vol. 17 No. 2, pp. 13-17.

J. Liang, C. Shaw, and M. Green (1991). "On Temporal-Spatial Realism in the Virtual Reality Environment." *Proceedings of the Fourth Annual Symposium on User Interface Software and Technology*, pp. 19-25.

M. A. Livingston, and A. State (1997). "Magnetic Tracker Calibration for Improved Augmented Reality Registration". *Presence: Teleoperators and Virtual Environments*, Vol. 6, No. 5. pp. 532-546.

D. Reiners (1999). "High-Quality High-Performance Rendering for Multi-Screen Projection Systems". *Proceedings of the 3rd International Immersive Projection Technology Workshop*, pp. 191-200.

R. Smith (1996). "Really Getting Into Your Work: The Use of Immersive Simulations". *Proceedings of the Symposium on Virtual Reality in Manufacturing Research and Education*.